\documentclass[acmlarge]{acmart}
\usepackage{booktabs} 
\usepackage{pdfpages}
\usepackage{graphicx}
\usepackage[ruled]{algorithm2e} 

\acmJournal{IMWUT}
\acmVolume{3}
\acmNumber{1}
\acmArticle{18}
\acmYear{2019}
\acmMonth{3}
\makeatletter
\let\@authorsaddresses\@empty
\makeatother
\setcopyright{rightsretained} 
\acmJournal{IMWUT}
\acmYear{2019} \acmVolume{3} \acmNumber{1} \acmArticle{18} \acmMonth{3} \acmPrice{}\acmDOI{10.1145/3314405}
\received{November 2018}
\received[Accepted]{January 2019}
\vbadness=\maxdimen
\begin{document}
\title{Animo: Sharing Biosignals on a Smartwatch for Lightweight Social Connection}
\author{Fannie Liu}
\email{ffl@cs.cmu.edu}
\affiliation{%
  \institution{Carnegie Mellon University}
}
\author{Mario Esparza}
\email{mesparza@snap.com}
\affiliation{%
  \institution{Snap Inc.}
}
\author{Maria Pavlovskaia}
\email{maria@snap.com}
\affiliation{%
  \institution{Snap Inc.}
}
\author{Geoff Kaufman}
\email{gfk@cs.cmu.edu}
\affiliation{%
  \institution{Carnegie Mellon University}
}
\author{Laura Dabbish}
\email{dabbish@cs.cmu.edu}
\affiliation{%
  \institution{Carnegie Mellon University}
}
\author{Andr\'es Monroy-Hern\'andez}
\email{amh@snap.com}
\affiliation{%
  \institution{Snap Inc.}
}
\renewcommand{\shortauthors}{F. Liu et al.}

\begin{abstract}
We present Animo, a smartwatch app that enables people to share and view each other's biosignals. We designed and engineered Animo to explore new ground for smartwatch-based biosignals social computing systems: identifying opportunities where these systems can support lightweight and mood-centric interactions. In our work we develop, explore, and evaluate several innovative features designed for dyadic communication of heart rate. We discuss the results of a two-week study (N=34), including new communication patterns participants engaged in, and outline the design landscape for communicating with biosignals on smartwatches.
\end{abstract}


\thanks{Authors' addresses: Human-Computer Interaction Institute, School of Computer Science, Carnegie Mellon University, 5000 Forbes Ave., Pittsburgh, Pennsylvania 15213; Snap Inc., Seattle, WA, 98121. Contact email: ffl@cs.cmu.edu}

\maketitle

\section{Introduction}
\begin{figure}[t!]
\centering
  \includegraphics[width=.4\columnwidth]{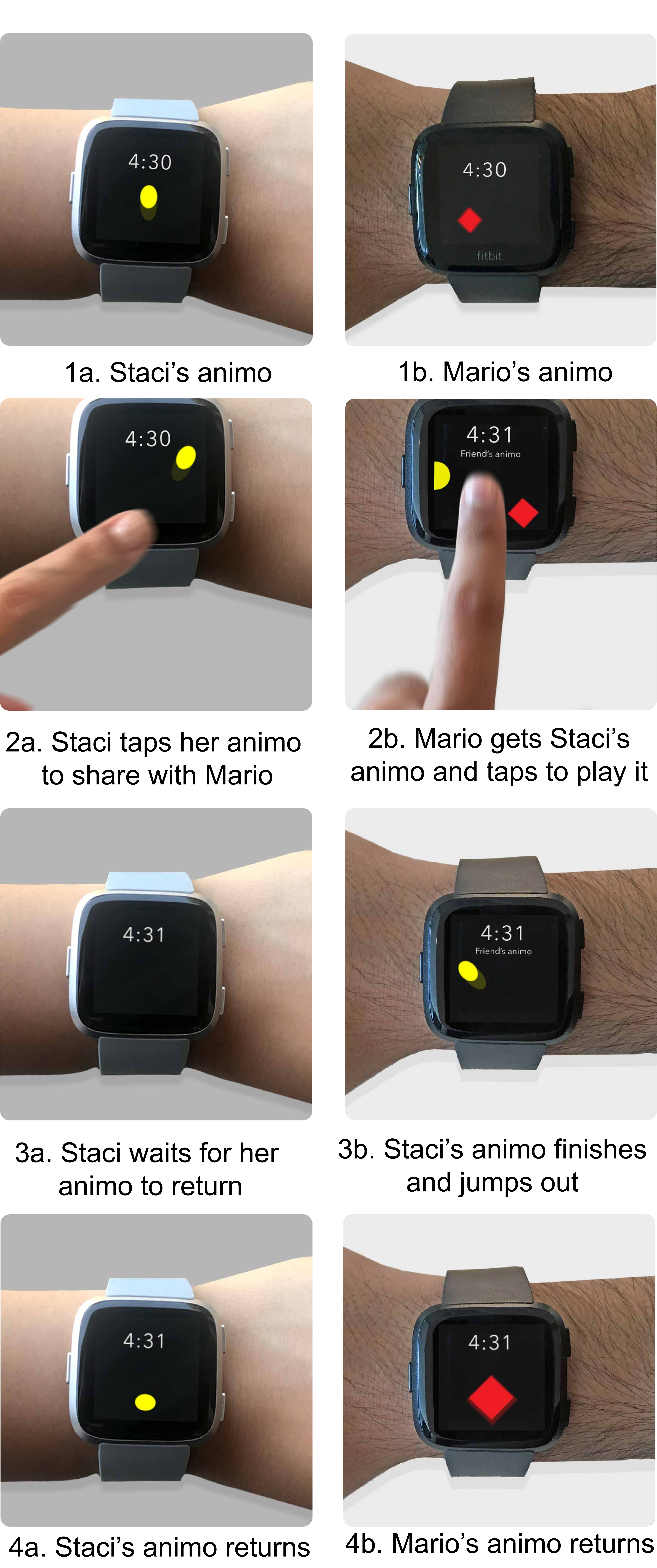}
  \caption{User experience of hypothetical dyad: Staci (left) and Mario (right) interacting via Animo.}~\label{fig:system}
\end{figure}

New forms of technology enable us to share a variety of personal data with each other--including our own heart rate. By wearing watches equipped with optical sensors, people can record and share their heart rate on their phones or computers while going for a run or streaming their gameplay~\cite{robinson2016all,curmi2013heartlink}. Recent research shows that sharing biosignals like our heart rate can provide social cues about our emotions or activities. By using these \textit{expressive biosignals} to convey our internal experiences, we can potentially become more aware and understanding of each other~\cite{liu2017supporting,howell2016biosignals,slovak2012understanding,hassib2017heartchat}.

While research suggests that expressive biosignals can facilitate interpersonal communication, integrating biosignals seamlessly into communication remains a challenge. Given their novelty as a cue, expressive biosignals face issues in interpretation, cognitive load, and privacy~\cite{liu2017expressive,liu2017supporting}. To address these issues and further explore the design space of expressive biosignals, we propose a new way to share biosignals: sharing them directly on smartwatches themselves.

Smartwatches could provide an unobtrusive and unique platform for sharing biosignals directly to another person. Many smartwatches already have built-in sensors that enable continuous monitoring of biosignals like heart rate (e.g., Apple Watch, Fitbit Versa, Mio SLICE, etc.), and thus do not require additional equipment to record the data. The form factor of a smartwatch could also afford intimate and vivid interpersonal communication. Being physically on the body, the device would be noticeable, easily accessible, and tangible--factors that can promote social presence and connectedness~\cite{van2015social,ijsselsteijn2003staying}. Moreover, exploring expressive biosignals on smartwatches would advance research on smartwatch communication, which suggests that beyond simply extending text or call notifications~\cite{jeong2017smartwatch}, the smartwatch itself could offer a lightweight yet rich communication channel~\cite{kim2016yo}.

Our research is motivated by two questions. First, how does sharing biosignals on a smartwatch impact communication? Specifically, what kinds of communication does this afford and what patterns emerge? Second, how do people make sense of their own biosignals and develop that understanding with each other? Given that biosignals are often ambiguous~\cite{howell2016biosignals,liu2017expressive} and the smartwatch screen has limited space, we aim to pinpoint the information necessary for meaningful communication.

We examine these questions through the design and deployment of a smartwatch app, and reflect on the design landscape that this work uncovers.

The main contributions of this work are:
\begin{itemize}
\item The design and implementation of Animo, a smartwatch app where people interact using biosignals.
\item A two-week user study of how 17 dyads used Animo and, consequently, interacted in novel ways.
\item An articulation of the design space for mood-centric social computing systems on smartwatches.
 \end{itemize}

\section{Related work}
\subsection{Communication on Smartwatches}
With the rising popularity of smartwatches, more researchers are exploring their capabilities. Given their ubiquitous nature and built-in sensors (e.g., biosensors, accelerometer, GPS, etc.), many works use smartwatches to understand and learn human behaviors, especially for health monitoring~\cite{amiri2017wearsense,vaizman2017recognizing,Capodieci2018}.
In a similar vein, research regarding the smartwatch user experience shows that personal monitoring activities, such as fitness and activity tracking, are some of the most popular features of the watch~\cite{watchusage,pizza2016smartwatch,jeong2017smartwatch}.

Fewer studies have explored the social applications of smartwatches. Mobile communication typically occurs via phone calls and text messaging, which are difficult on a small screen on the wrist. Subsequently, many researchers are exploring better text entry for small-screen devices~\cite{textentry1,textentry2,textentry3,textentry4,textentry5}. However, we argue that smartwatches enable opportunities for communication less focused on text. 
In fact, much of communication online contains non-textual cues. Emojis are the most common, and are used for emotional information or for expressive and playful interactions~\cite{derks2008role,zhou2017goodbye,wiseman2018repurposing}. People also share non-textual cues for their context, such as their location~\cite{tang2010rethinking,smith2005social,massa2015if} or activity~\cite{epstein2015nobody,rooksby2014personal}. Even ``one-click communication,'' such as ``liking'' online social media posts, can provide diverse cues (e.g., support, agreement)~\cite{scissors2016s}.

Smartwatches have the capability of communicating in a non-textual way. Kim and colleagues provide one example: using the Yo app as inspiration, they suggest conveying affect or location through the smartwatch. Using simple yet expressive imagery (e.g., kinetic typography) or built-in sensors (e.g., GPS), they describe the potential for rich single touch messaging on the smartwatch~\cite{kim2016yo}. In the present work, we build on this research by exploring the non-textual communicative abilities of the smartwatch, using biosignals.

Biosignals present an opportunity to explore novel and expressive communication cues afforded by the smartwatch. Most existing non-textual cues are easily accessible on platforms other than the smartwatch, such as smartphones or desktop computers. Biosignals, on the other hand, are more easily and unobtrusively accessed on the smartwatch than on other platforms. For example, smartphone apps that record heart rate require users to place their finger on the phone's camera for measurement, while smartwatches with heart rate sensors can record heart rate passively and continuously, only requiring users to wear the watch. Moreover, while emojis are the most common non-textual cue, there are hundreds of emojis with continuously evolving definitions~\cite{lu2016learning,miller2016blissfully,wiseman2018repurposing}, which increases the cost of communication when selecting one on a watch. Biosignals represent our body's immediate response to situations, and thus could provide more authentic and expressive cue by capturing the body's underlying state at specific moments. In the following section, we describe research investigating the potential for biosignals to act as a social cue.

\subsection{Sharing Biosignals}
Like smartwatches, most prior work in biosignals target individuals, such as providing users with personal feedback on fitness, well-being, or social skills~\cite{frey2018breeze,rubin2015towards,pancardo2016personalizing,griffiths2014health,anderson2013tardis,konstantinidis2009using}. However, some research has explored the potential for biosignals to support interpersonal communication, focusing on how they affect our emotions~\cite{kreibig2010autonomic,posner2005circumplex}. For instance, researchers have built chat systems that utilize biosignals like brain activity to detect affect, and convey how a user is feeling through avatars~\cite{kuber2013augmenting} or kinetic typography~\cite{wang2004communicating}. More recent work has explored expressing feelings through displays of biosignals themselves. \textit{EmpaTalk} and \textit{HeartChat} are systems that include visual indicators of heart rate changes (e.g., graphs, colors, raw beats per minute) to help online chat partners communicate their emotions~\cite{lee2014empa,hassib2017heartchat}. Tan and colleagues developed a system that displayed changes in heart rate, skin conductance, and respiration in order to reduce stress in worker/instructor-based collaboration~\cite{tan2014investigating}. Researchers have labeled these displays as \textit{expressive biosignals}, in which biosignals are used as social cues about someone's emotional and/or cognitive states~\cite{liu2017supporting,liu2017expressive}.

A few studies have explored how expressive biosignals can affect social interactions. Janssen and colleagues, for instance, demonstrate that the sound of a heartbeat can increase feelings of intimacy and closeness~\cite{janssen2010intimate}. Sharing heart rate can increase social connectedness, where it can become an emotional expression, a means to gain awareness about another person's context, or simply a playful way to interact and feel present with another person~\cite{liu2017supporting,hassib2017heartchat,slovak2012understanding}.

Most expressive biosignals systems are built as smartphone or desktop apps. Only a few systems exist on wearables, such as a helmet~\cite{walmink2014displaying} or shirt~\cite{howell2016biosignals,howell2018tensions}. These systems focus on broadcasting the data to all viewers. On the other hand, smartwatches have a smaller form that could enable direct communication and can also be found with built-in biosignals sensors. For example, Apple Watch's Digital Touch app allows users to directly share vibrations of their heart beats to another person. However, the Digital Touch does not monitor heart rate continuously, and thus requires users to spontaneously think about sharing with someone else, which is unlikely to happen given the unconventional nature of sharing heart beats. Thus, to understand and explore the design of expressive biosignals smartwatch systems that encourage sharing, we created our own app: Animo.

\section{Animo system}
\subsection{Design}
Animo is a smartwatch app where two people can send mood representations, or ``animos'' (lower case), to each other. We referenced existing frameworks for augmented mobile messaging systems~\cite{buschek2018personal,hassib2017heartchat} to inform the design of Animo. Though this prior work focused on augmenting text messaging for the phone, many of the same concepts can apply to non-textual smartwatch communication.

\subsubsection{Content from Sender}

Buschek and colleagues' design space for augmented mobile messaging includes the ``Sender Context'' dimension, describing context as ``information and cues beyond text''~\cite{buschek2018personal} to be included with a text message. Animo does not use text; therefore, we describe its communication~\textit{content}. Since our focus is biosignals, heart rate is the content type. Heart rate sensors are a common feature of many smartwatches, and people already associate heart rate
with different psychological states, such as mood~\cite{slovak2012understanding}.

Previous studies demonstrate that simply showing heart rate as a raw value can limit expressiveness and appeal, and can be difficult to understand~\cite{liu2017supporting,buschek2018personal}. Thus, inspired by the popularity of mood rings, we chose to represent heart rate as ``mood,'' to guide engagement and understanding.

The \textit{system} is the \textit{content provider}, where Animo shows a mood representation (animo) to a user; a user cannot choose the animo. We base mood loosely on the valence-arousal circumplex, which separates emotion into two dimensions: valence (positive/negative) and physiological arousal (high/low)~\cite{posner2005circumplex}. Given the constraints of information available on smartwatches, we focused on mood related to physiological arousal determined by users' heart rate (e.g., excitement as a high-arousal mood vs. calmness as a low-arousal mood), and left valence open to users' interpretations.

\subsubsection{Sharing}
Users can view their animo on their smartwatch, and tap on it to send it to their partner. Their partner, who must also be running Animo, will feel a subtle vibration on their watch when they receive the animo. The animo  then ``peeks'' into the side of their watch screen. Tapping on it will play its animation. Given privacy concerns around sharing biosignals~\cite{liu2017supporting,liu2017expressive}, users can \textit{explicitly} and \textit{sporadically} share their animos as they please. To encourage sharing, the watch occasionally vibrates when their animo state changes.

\subsubsection{Presentation Abstraction} We chose a \textit{high} abstraction representation for mood derived from heart rate. To enhance expressiveness and playfulness, animos are animated shapes.

We designed different animos to cover a variety of moods. 
For some designs, we drew from elements of kinetic typography that were tested in Kim and colleague's work on  Yo~\cite{kim2016yo}. The animos varied in three ways:

\begin{itemize}
\item \textbf{Shape:} In a dyad, one person has circle animos while the other has diamond animos. Shapes are assigned during onboarding and do not change.

\item \textbf{Motion:} We designed animos with different levels of energy to represent arousal. High energy motions (e.g., bouncing) represent higher heart rate; low energy motions (e.g., swaying) represent lower heart rate.

\item \textbf{Color:} As with mood rings, some degree of mystery can encourage playful discovery. Therefore, we designed animos to change colors semi-randomly in order to encourage users (who would not be aware of the randomness) to question and interpret their animos. High energy animos are randomly yellow or red, whereas low energy animos are randomly blue or green. Animos in between high and low energy levels are white. We chose these colors according to their existing associations with emotions~\cite{colors}, loosely basing them on their relation to the valence dimension of mood but still leaving them up to interpretation.
\end{itemize}

We pretested the animos we designed on the crowdsourcing platform Amazon Mechanical Turk to ensure general agreement that they represented expected moods and their associated arousal levels. We tested a total of 26 different animos across three rounds of surveys\footnote{We ran multiple rounds of surveys in order to collect a diverse set of animos that would cover the quadrants of the valence-arousal circumplex. In between rounds, we created new animos when we did not have enough animos that performed well in certain quadrants (e.g., designing for the positive/low arousal quadrant was particularly challenging)}. In each round, 20 participants viewed a subset of animos presented in a randomized counter-balanced order and rated them on their ``mood'' and ``energy''~\cite{lee2006using} (see supplemental materials for the survey questions). We selected the best performing animos per round to include in the Animo app, which led to a total of 18 animos (see supplemental materials for the pretest analysis and results, and a video of the selected animos). Four sets of three animos were chosen to cover different quadrants of the valence-arousal circumplex, where they differed significantly in mood and energy ratings ($p \leq 0.05$). We also included two animos that differed significantly in only energy ratings, in order to introduce some ambiguity that could spark different interpretations. Finally, we included four ``neutral'' animos that were not significantly high or low in mood or energy ratings.

\subsubsection{Presentation Granularity}
Animo is \textit{person-based}, meaning each user has their own animo. Users can view their own animo on their smartwatch, and their partner's if they send it to them. Users can have one, and only one, partner. We made this decision to focus on the simplest communication on a smartwatch---one-to-one. Animo is also \textit{message-based}: each sent animo is based on a user's current state.

\subsubsection{Presentation Persistence}
Sent animos are \textit{ephemeral}: they disappear in 10 seconds if the receiver ignores them by not tapping on them. This emphasizes animos' weightlessness, and aims to avoid having yet another feed to check. We chose 10 seconds based on early pilot tests we conducted to ensure that receivers would have enough time to notice and tap the animo if desired.

\subsection{Implementation}

We implemented Animo as a ``clock face'' app for the Fitbit Versa smartwatch, which allowed it to stay on the default screen of the smartwatch. We chose the Fitbit platform because of its compatibility with both Android and iOS, and its ``mass appeal,'' with over 25 million active users in 2017~\cite{fitbitnum, massappeal}. The Fitbit Versa has an LCD touchscreen, Bluetooth communication, and a heart rate sensor, among other sensors.

\begin{figure}[t!]
\centering
  \includegraphics[width=.7\columnwidth]{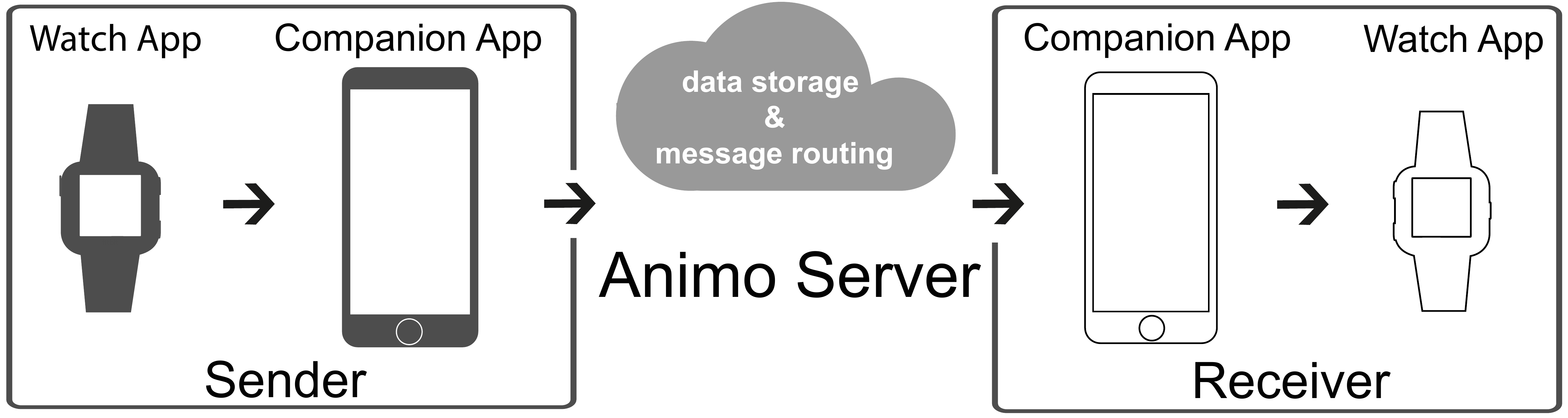}
  \caption{Components of the Animo system.}
  ~\label{fig:architecture}
\end{figure}

The Animo system is composed of a smartwatch app, a server-side app running on the cloud, and a ``companion app" that runs within the Fitbit smartphone app (see Figure ~\ref{fig:architecture}). The companion app enables communication between the Fitbit Versa and a user's smartphone via Bluetooth, which is used when a user sends an animo.

The companion app sends the animo to the backend service, which routes the animo to the user's partner.
This alerts the receiver's smartwatch that an animo is available, triggering a vibration. The animations are displayed only on the smartwatch and implemented using vector graphics. 
Animo requires the sender and receiver to both have the Fitbit Versa and the Fitbit app running on their phone with Bluetooth and data connection on.

\section{Methods}
To test Animo \textit{in situ}, we deployed the app in a two-week field study, allowing participants to freely use the app in order to observe patterns of usage that naturally emerge. 

\subsection{Participants}

\begin{table}[tbp]
\centering
\footnotesize
\caption{Participant dyads. Includes drop outs ($\dagger$). Some friends (*) reported also being coworkers.}
\label{tab:participants}
\begin{tabular}{@{}llrrlrrr@{}}
\toprule
\multicolumn{5}{c}{\textbf{dyads}} & \multicolumn{3}{c}{\textbf{animos}} \\ \midrule
\multicolumn{2}{l}{\textbf{participants}} & \multicolumn{2}{r}{\textbf{genders}} & \textbf{relationship} & \textbf{sent} & \textbf{read (\%)} & \textbf{replied (\%)} \\ \midrule
P1$\dagger$ & P2$\dagger$ & M & M & friends* & --- & --- & ---\\
P3 & P4 & M & M & friends* & 220 & 40 (18\%) & 18 (8\%) \\
P5 & P6 & F & F & friends* & 127 & 65 (51\%) & 25 (20\%) \\
P7$\dagger$ & P8$\dagger$ & F & F & friends & --- & --- \\
P9 & P10 & M & M & friends & 175 & 115 (66\%) & 59 (34\%)\\
P11 & P12 & F & F & friends* & 173 & 101 (58\%) & 43 (25\%)\\
P13 & P14 & M & F & significant others & 258 & 77 (30\%) & 66 (26\%) \\
P15 & P16 & M & F & spouses & 68 & 34 (50\%) & 7 (10\%)\\
P17 & P18 & F & M & significant others & 45 & 14 (31\%) & 6 (17\%)\\
P19$\dagger$ & P20$\dagger$ & F & F & friends & --- & --- & ---\\
P21 & P22 & M & F & significant others & 210 & 113 (54\%) & 60 (29\%)\\
P23 & P24 & M & F & spouses & 77 & 35 (45\%) & 16 (21\%) \\
P25 & P26 & F & F & roommates & 108 & 43 (40\%) & 6 (6\%) \\
P27 & P28 & F & M & spouses & 33 & 3 (9\%) & 2 (6\%)\\
P31 & P32 & F & F & friends* & 168 & 87 (52\%) & 39 (23\%) \\
P33 & P34 & M & M & coworkers & 375 & 159 (42\%) & 35 (9\%) \\
P35 & P36 & M & F & spouses & 181 & 43 (24\%) & 25 (14\%) \\
P37 & P38 & M & F & spouses & 115 & 49 (43\%) & 29 (25\%) \\
P39 & P40 & F & F & roommates & 90 & 27 (30\%) & 8 (9\%) \\
P41 & P42 & F & M & coworkers & 67 & 35 (52\%)  & 4 (6\%) \\ \bottomrule
\end{tabular}
\end{table}

We recruited 20 dyads, or 40 participants. We removed data from three dyads, leaving a total of 17 dyads (see Table ~\ref{tab:participants}).
Participants were removed either because they experienced major technical issues (e.g., loss of connection between their phone and the smartwatch), or because they were traveling without connectivity for a majority of the study.

We recruited participants through the mailing lists of a technology company, inviting people to participate in a two-week experiment about ``mood.'' 
We did not pay participants. We asked participants to choose a partner to join them in the study.
This partner did not have to be affiliated with the company. In order to have a diverse sample, we recruited participants from three different offices in the United States: New York City, Seattle, and Los Angeles.

Participants varied in their backgrounds and demographics. Their occupations included homemaker, professional server, program manager, business recruiter, software engineer, neuroscientist, and others.
 Their ages ranged from 19 to 48 years old (\textit{M}\textsubscript{age} = 30.4 years, \textit{SD}\textsubscript{age} = 6.0 years). Participants' gender and relationship breakdown can be seen in Table~\ref{tab:participants}. 
Sixteen participants identified as Asian, 12 as White/Caucasian, two as Hispanic, and three as mixed White/Caucasian and Asian or White/Caucasian and Hispanic. Fifteen participants owned a smartwatch (e.g., Apple Watch, Google Wear OS), with nine of them using it frequently for activity tracking, phone notifications, or heart rate monitoring. Participants' prior usage of smartwatches did not influence our results, therefore we included all participants who owned smartwatches in our final analyses.

\subsection{Procedure}

\subsubsection{Onboarding} Each dyad was onboarded together in one of the offices of the technology company. Participants first created a Fitbit account and added each other as friends on Fitbit, then individually completed a questionnaire to describe their backgrounds (see supplemental materials for the questionnaire).

Next, experimenters equipped participants with a smartwatch, and took their heart rates during a calming task (individually watch a breathing exercise video\footnote{Video available at \url{https://www.youtube.com/watch?v=5f5N6YFjvVc}}) and during a stressful task (count down from 1022 in steps of 13~\cite{birkett2011trier} in front of their partner and the experimenters).
We used the average of the heart rates recorded during each task to determine animo arousal, i.e., high and low heart rate baselines.

After recording the heart rates, experimenters explained to participants how to use Animo, including how to send and view received animos. Experimenters purposefully \textit{did not} explain the meaning behind the different animos, and instead instructed participants to interpret the animos themselves. We made this decision to inform our research questions, allowing participants to flexibly define the animos to understand how they would create those definitions. Additionally, prior work highlights the importance of allowing people to create meaning together from their biosignals~\cite{liu2017supporting}. Finally, participants could leave and use Animo freely for two weeks.

\subsubsection{Animo Usage}
We recorded a variety of data to capture participants' Animo usage throughout the study.
This included heart rate data, animo states, and animos sent, received, viewed, and sent as responses. Additionally, inspired by diary studies~\cite{diarystudy}, we sent brief daily surveys about their usage (see supplemental materials for the survey). 
After one week, participants completed a mid-study survey to clarify responses in their daily surveys and provide initial thoughts on Animo.

\subsubsection{Offboarding}
After two weeks, participants returned their watches and were individually interviewed. 

The interview was semi-structured and elicited participants' thoughts and feedback on their experiences with Animo (see supplemental materials). 

To help participants recall their experiences, we showed them the five animos they sent the most and the five animos received the most.

In the feedback section, we also showed example sketches and mock-ups for future Animo designs, to probe participants on specific aspects of Animo that they enjoyed or wanted to improve (e.g., what animos looked like, how often they should be sent).

\subsection{Analysis}
We analyzed the responses to the daily surveys, mid-study surveys, and transcriptions of the audio-recorded exit interviews, using participants as the unit of analysis. Two researchers independently performed open-coding to label responses from a random sample of participants. They developed codes according to similarities in participants' overall usage of Animo, experiences sending/receiving animos and subsequent reactions, process of understanding the animos, and feedback for Animo. The researchers met frequently to discuss these codes and create a codebook. Once they agreed on the codebook, one researcher used the codebook to code the rest of the participants. Next, we grouped related codes together and formed themes around participants' communication patterns and how their understanding of the animos affected those patterns and their attitudes towards Animo. During the writing process of the paper, we refined the themes around our main research questions. 

\section{Results}
Participants used Animo frequently throughout the study, averaging \emph{five animos sent per user per day} despite not being required to do so and not receiving compensation for participating.
Four participants even continued using Animo after completing the study. Across all participants, a total of 2,490 animos were sent, and 1,040 (41\%) were read (participants received their partner's animo, and tapped on it to view the animation). Of the animos read, 43\% received a reply on average (participants sent an animo back within 10 minutes after reading one). Of the animos sent, 5.6\% were lost due to connectivity issues, such as unstable Bluetooth connection between users' phones and the smartwatch or participants traveling to areas with limited network connectivity.

In the following, we describe our results from participants' responses to the surveys and interviews, providing a richer view into the reasons behind participants' Animo usage. We detail the common themes that emerged in our results, and within those themes, include interesting examples of Animo usage that participants shared. 

\subsection{Connecting in New Moments}

Animo's design allowed participants to more easily stay connected with each other. Its convenience and constant physical access afforded communication when they typically did not or were not able to communicate with their partner.

\subsubsection{Seeing Animo is a Reminder to Communicate}
Participants found that looking at their watch reminded them of their partner, and made their partner more \textit{salient}. Since the smartwatch is easily accessible on their wrist, participants only needed to glance down to see their animo. They found themselves doing so not only by haptic prompting (i.e., the watch vibrating when the animo state changes or when receiving an animo), but when they were bored, had ``down time,'' or simply wanted to check the time.

For instance, P22 was traveling in a different city than P21, her significant other, during the first week of the study. She noted that while she was away from P21, the \textit{``watch represented a connection to [him]''}, and \textit{``increased communication when there wouldn't normally have been communication''}:

\begin{quote}
\textit{``...because, like,the thing on my wrist was [him]. It, like, reminded me that, like, there was a prompt to communicate.''} ---P22
\end{quote}

Looking at Animo thus prompted participants to think of their partner, and sending animos let their partner know that. As a result, participants described feeling happy and nice whenever they received their partner's animo:

\begin{quote}
\textit{``It's, like, kind of like getting a `like' on Twitter or on Instagram. You're, like, `Oh, somebody thought about me!' And they're not really thinking about you, but they're trying to, like, show you your existence. I like that you're there.''} ---P11
\end{quote}

\subsubsection{Communication is Convenient through Animo}
Participants also used Animo when they were already thinking of their partner but unable to communicate through other means. Tapping on their watch was more convenient than \textit{``patiently typing a message''} (P24) on the phone or \textit{``[having] a computer ready to go''} (P15). Animo sending occurred frequently when participants were too busy attending to something else to communicate otherwise. For example, even when P22 was busy traveling, her partner (P21) felt he could easily keep in touch:

\begin{quote}
\textit{``I think she was out with her... parents and not necessarily able to text back and forth, but you could still send each other animos...''} ---P21
\end{quote}

\begin{figure}[t!]
\centering
  \includegraphics[width=0.8\columnwidth]{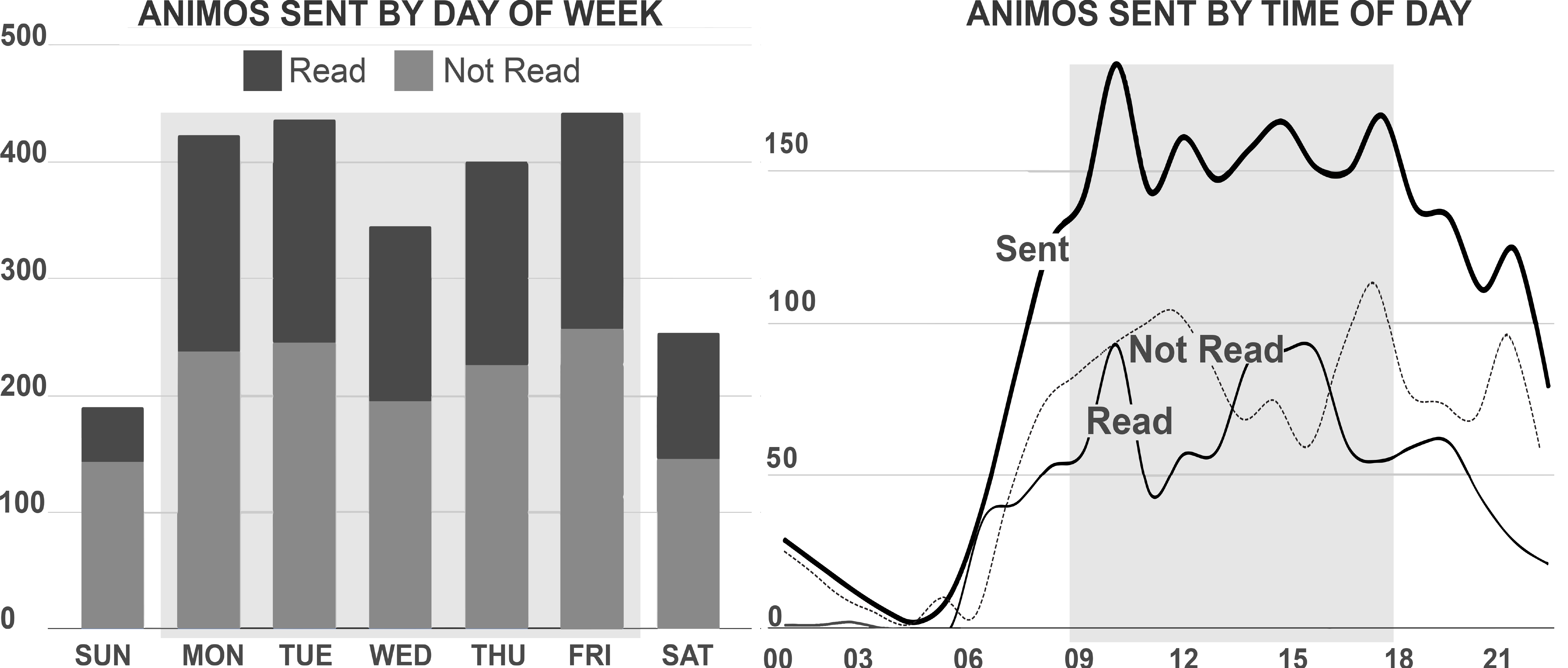}
  \caption{Usage patterns by time of day. Working hours are highlighted in gray.}~\label{fig:usage}
\end{figure}

Similarly, participants felt that they were able to send animos during their busy work day, when they typically were not physically co-located with their partners. Our data on Animo usage supports this (Figure \ref{fig:usage}), with the highest number of animos sent and read during work days (Monday-Friday) and work hours (9am-6pm).

Since tapping on a watch is less noticeable than taking out a phone or speaking, Animo provided a private communication channel while in public. For example, when P5 and P6 were in a stressful work meeting together with other colleagues, P6 used Animo to communicate with P5 whenever she thought they would both be annoyed. She described sending animos to let P5 know she was \textit{``thinking about her without being obvious in [the] meeting.''}

Our results suggest that Animo may enhance the salience of partners and increase communication. Animo's unobtrusiveness and ease of interaction on the smartwatch allowed participants to connect in moments they were not together, were too busy, or needed privacy in a public space.

\subsection{Creating New Understanding}
Through Animo, participants felt they could elicit new information from their partners about their states.
Receiving animos gave participants insight on their partner's state and opened up new pathways for them to discuss and understand each other's state. This insight aligns with prior research on sharing heart rate~\cite{liu2017supporting,hassib2017heartchat}, where participants felt that shared biosignals functioned as an emotional expression or status update. Our work extends this prior work by detailing how status awareness develops through shared biosignals.

\subsubsection{Animo can Start Conversations about Your Day}
Animo triggered participants to start \textit{new conversations} with each other, where they checked in with each other based on the animos they saw throughout the day. For example, P13 and P14, who were in a long-distance relationship and only saw each other every few weeks, used animos as a conversation starter while they were apart:

\begin{quote}
\textit{``She'd get one that was white. She's like, 'You're relaxing right now?' I said, 'Yeah, I'm at home right now.' I was just reading and it prompted other conversation...to check in with each other and see what we're up to.''} ---P13
\end{quote}

Even when participants were too busy to have these kinds of conversations during the day, the animos provided a \textit{``tapestry''} or \textit{``glimpse''} (P18) of their day. Participants then talked about the animos with their partners when they later saw them face-to-face. This was especially common for participants that lived together (i.e., significant others and roommates). For example, P25 and P26 were roommates who usually briefed each other on the upcoming events of their day before they left for work. P25 once sent an animo to express that she was stressed before an interview. P26 described following-up on that animo:

\begin{quote}
\textit{``[W]hen we saw each other in the evening, uh, we would like mention that we had sent them...so I like asked if she was in the interview when she sent it to me so we talked about that. So it just kind of prompted discussions about our day.''} ---P26
\end{quote}

When participants received an animo that triggered their concern, they would start conversations immediately. This occurred when participants thought their partner was stressed. For example, P14 learned that her partner had a frustrating experience with his car. Soon after hearing about it, she received a red animo (which she interpreted as stress), and called her partner immediately because she was worried.

While Animo helped start conversations, participants felt it could not support full conversations on its own. For instance, when participants received animos, sending an animo back only represented an acknowledgement of receipt. P15 felt that Animo acted as a \textit{``first level of communication,''} to start conversation, and desired a way to easily access or inform the next level of communication, where conversations could actually take place:

\begin{quote}
\textit{``[If] I see this yellow jumping dot, it probably means that you know my wife wants to talk about something fun and then oh let me try to call her. So, if there was something, like, oh whenever I share this I want my wife to be able to call me back.... If there was some way to establish this...second level communication would be nice.''} ---P15
\end{quote}

\subsubsection{Animo can Clarify Ambiguous Conversations}
In a less common yet interesting case, Animo helped participants understand their partner's feelings when they had difficulties interpreting their behavior. P42, who had a coworker/work-friend relationship with P41, used Animo to validate his thoughts on how P41 felt. After an in-person conversation with P41 where she seemed \textit{``riled up,''} P42 checked the animos she sent:

\begin{quote}
\textit{``It was interesting to see if...one of us was more upset about something...it was good to know if this person was actually worked up or if this was a show, like a front.''} ---P42
\end{quote}

Though P42's experience was a unique example in our results, it reveals Animo's potential to prompt deeper understanding between people by starting new kinds of conversations and clarifying ambiguous feelings. This finding supports and extends results from prior research showing that shared heart rate can provide emotional status cues and open up communication about those cues~\cite{liu2017supporting}.

\subsection{Navigating Open-Ended Interpretation}
The Animo system provides an abstracted representation of sensed mood. Given the level of ambiguity we imbued in the animos (e.g., random colors) and the lack of provided definitions, participants' understanding of that abstraction varied.

We found that participants' interpretations did not necessarily align with the results of our pretest of the animos--instead, participants situated animos in social contexts. Participants' interpretations affected whether they found the animos \textit{meaningful}, and determined their engagement with Animo. Below, we detail how Animo's open-ended nature impacted participants' ability to meaningfully communicate.

\subsubsection{Playful Imagination and Discussion}
Participants enjoyed having free reign to understand the animos for themselves. In particular, they appreciated that Animo did not necessarily tell them their mood, but encouraged them to reflect on their mood themselves. P15 compared this to emojis:

\begin{quote}
\textit{``I like that animos...looked unintentional in a way that they don't necessarily imply something you know, whereas like an emoji definitely implies something. It doesn't make me feel I'm being forced in feeling something in some way.''} ---P15
\end{quote}

Participants found it fun to not only think about how their own animo could relate to their mood, but also imagine what their partner might be doing, and why they chose to send the animo they did. For example, P21 felt that because the animos were not always accurate, there was more meaning to the ones they selected:

\begin{quote}
\textit{``It's just fun to receive them because...she looked at it and then she thought that that reflected her mood and then chose to send it. So, it feels meaningful because of that...she chose that one specifically.''} ---P21
\end{quote}

Participants also enjoyed the process of decoding the animos with their partners. They noted that they became more aware of each other's feelings, \textit{``not just with the animos but...in talking to each other''} (P6). Even seeing less accurate animos spurred these discussions:

\begin{quote}
\textit{``I mean, like, sometimes we were just, like, laughing about it. I think, like, it allowed us to kind of just have conversations based on our feelings and how we were feeling in the moment.''} ---P37
\end{quote}

\subsubsection{Finding Mismatched Meanings}
Though most participants found meaning in the animos, where they felt that their animos reflected or somewhat reflected their state, the meaning they gleaned did not always match the meaning their partner gleaned.

\paragraph{Missing a shared language.} Participants were not always able to have conversations to jointly reflect on animos with their partner. When this happened, they would be unable to determine whether they agreed on the meanings they attributed to the animos. Instead, they tended to reflect according to their own beliefs.

For example, P11 and P12 were close friends and coworkers who would send animos to each other both as a ``poke'' and to show that something interesting was happening. P11, who would send her animo to let P12 know what her mood was, stated,  \textit{``I wanted her to know that it was matching my mood but, um, I don't think she knew.''} Indeed, P12, who believed the animos were more related to physical activity rather than mood, instead reflected on P11's activity:

\begin{quote}
\textit{``So, like, for the white ones... she's probably just lying in bed or sitting around somewhere doing something. For active ones.. .maybe she's like jumping around or dancing or, like, walking about or exercising when she sent it.''} ---P12
\end{quote}

\paragraph{Unintended interpretations.} Diverging opinions on the meaning of the animos sometimes led to unintended interpretations. For example, P42 viewed animos as representative of his mood--to the extent that he felt Animo was making him more aware of it. He similarly believed P41's animos represented her mood, and thought she sent her animos to show him her mood. In actuality, P41 did not find meaning in the animos and sent them at random times. This suggests that an animo could reveal more information than a sender expects (recall that P42 used Animo to validate his thoughts on P41's feelings), and that a receiver can interpret an animo in ways that stray from a sender's intentions.

\paragraph{Giving animos context.} Some participants realized their partner might interpret their animos differently than intended. To counter this, they would send additional information through other communication channels to clarify animos they sent. For example, P38 sent his partner a red animo to show that he is excited. However, he recognized that red could be interpreted as an angry color; therefore, he sent his partner a video to give them more clarifying context.

\subsubsection{Animo as Content-Less Notification}

Ten participants could not find meaning in the animos, and subsequently felt Animo had limited communicative ability. These participants sent animos to simply connect and say ``hi'' to their partner.

Dyads who believed their animos lacked meaning quickly got bored of Animo and used it less. P39 and P40, for instance, were roommates who both felt frustrated with the system because it made no sense to them and did not seem to match their how they felt (e.g., seeing red animos they view as angry when they actually feel happy). They would send their animo even when it did not match their feelings because their partner \textit{``wouldn't care, wouldn't realize, or wouldn't read into it''} (P39). They instead sent animos randomly as a ``hi.'' Communicating in this way had limitations, as there are only so many times people will say ``hi'' back and forth:

\begin{quote}
\textit{``...I just did not want to keep it going. I was like uh, `hi,' `hi', and `hi' is fine.''} ---P39
\end{quote}

Overall, we found that Animo's open-endedness had benefits and tradeoffs. Participants could decide what the animos meant to them, and reflect on those meanings as part of new conversations. However, not all participants were able to interpret animos, and even when they did, their interpretations did not always match their partner's without a clarifying conversation. This suggests a need to convey the intention of messages sent in systems like Animo, while maintaining the value brought by their brevity.

\section{Discussion}
Our results show that Animo enabled \textit{lightweight social connection}, where participants found it fun and easy to keep in touch and attuned to each other's presence and state. However, Animo experienced some challenges in functioning as a full communication platform, due to its minimal and somewhat ambiguous nature. Repurposing and expanding Bushek and colleagues' design space for augmented mobile messaging~\cite{buschek2018personal}, we discuss design implications for biosignals smartwatch communication systems based on opportunities and challenges we saw in Animo features. We summarize recommendations for this new design space in Table~\ref{tab:designrec}.

\begin{table*}[tbp]
\centering
\footnotesize
\caption{Recommendations for the biosignals smartwatch communication design space. Adapted from~\cite{buschek2018personal}.~\label{tab:designrec}}
\def\arraystretch{1.3}
\begin{tabular*}{\textwidth}{@{\extracolsep{\fill}} p{1.5cm}p{1.8cm}p{5.65cm}p{5.75cm}}
\toprule
\multicolumn{3}{l}{\textbf{Insights on existing design dimensions}} \\ \midrule
\multicolumn{2}{l}{\textbf{Dimension}} & \textbf{Recommendation} & \textbf{Support from Animo} \\ \midrule
Content & Provider: \newline {\tiny System | User | Mixed } & Combine system- and user-provided content, to help users create their own meaning from their biosignals rather than solely relying on the system. & Developing meaning for animos gave users creative liberty rather than being ``forced'' into feeling a certain way based on what the system told them. \\ \hline
Presentation & Abstraction: \newline{\tiny Low | Med | High} & For limited content,
explore expressive yet simple representations that contain clear and distinct information. & Expressing oneself through the open-ended animos was fun, but also inhibited developing a shared language because they could mean different things.\\ \cline{2-4}
 & Granularity: \newline {\tiny Person | Message}\newline {\tiny | Communication} & 
Communication-based granularity may promote playfulness, connectedness, and shared meaning. Example: messages that jump around each other when communication is exciting. 
 & Animos were cute and playful, but primarily represented individual users' moods, which users had to convey to their partner. \\ \toprule
\multicolumn{3}{l}{\textbf{Suggested new design dimensions}} \\ \midrule
Sharing & Receiver: \newline {\tiny One | Many} & 
Single-receiver is more intimate,
and watch becomes a reminder of that person. Multi-receiver allows
for keeping in touch with more people. & Having one partner made animos feel like a personal and private way to communicate. Some mentioned wanting multiple partners to check in on. \\ \hline
Response & Richness: \newline {\tiny Simple | Rich} &
Explore 
richer opportunities for receiver responses, beyond just sending a message back. Example: simple response could be an acknowledgement of message receipt; richer response could be in-app short text or animation. & Just sending an animo back was limited. Participants wanted more unique and richer responses that are inherently tied to the original sent message. \\ \cline{2-4}
 & Channel: \newline{\tiny Single | Multiple } & 
Link the watch to other channels for deeper conversation. Watch can act as a lightweight first level of communication, e.g., initiating conversation or simply ``being there.''
 & Since Animo communication was lightweight, participants typically followed-up in later conversations in-person, through text, or on the phone. Participants wanted a way to more smoothly enter those conversations directly from the watch. \\ \bottomrule
\end{tabular*}
\end{table*}

\subsection{Content and Presentation: Being Expressive yet Interpretable}
Bushek and colleagues describe a design space for augmenting text messaging with different types of context~\cite{buschek2018personal}. As a related but tangential form of messaging, biosignals smartwatch communication focuses on biosignals as the content of the message itself. Our results show that Animo promoted lightweight social connection through the \textit{content provider} and its \textit{presentation} to users, where some randomness and abstractness allowed users to have fun creating and discussing their own meanings together. At the same time, being too abstract and ambiguous can lead to challenges in finding meaning in the content.

\subsubsection{Expressive Meaning-Making}
Animo enabled participants to express themselves, even though the system intentionally provided them with a partially random mood representation. Participants found the process of interpretation fun and meaningful, both when they saw animos that did and did not reflect their mood. The ambiguity allowed them to be expressive by developing their \textit{``own vernacular''} (P27), as opposed to using pre-existing ones. This supports recent research on approaches to emotional biosensing, which questions systems that use biosignals in attempt to determine and tell people how they feel according to predefined emotional categories~\cite{howell2018emotional}, and suggests potential for people to have meaningful discussions around biosignals instead~\cite{liu2017supporting}. In the same vein, our results demonstrate the importance of enabling people to reflect on their biosignals and create their own meanings for them, rather than a system simply telling them what they mean. Additionally, the playfulness and expressiveness of Animo supported \textit{collective} meaning-making experiences for participants and their partners, which can ultimately facilitate communication and social connectedness ~\cite{hsieh2017playfulness,xu2017emphasizing}.

\subsubsection{Limitations of Abstraction}
While participants enjoyed being expressive through the highly abstract animo designs, animos were sometimes difficult to interpret. We provided animos as the sole content to reinforce Animo's lightweight nature; however, when that content was not understandable, communication became meaningless. Prior work suggests that communication that lacks content is limited: while it can contain a ``symbolic'' message (e.g., ``hi''), it has less value and impact on relationships~\cite{burke2014growing}. Based on our results, two factors appeared to affect the meaning of the content:

\paragraph{Cognitive effort} Decoding the information provided within the animos required more cognitive effort than expected. Though we had designed animos to convey mood through color and motion, most participants focused on only color. Using both color and motion to track animo states proved difficult when participants saw only one state at a time. Subsequently, they often assigned meaning to animos according to only one dimension of the valence-arousal circumplex~\cite{posner2005circumplex} (e.g., red animos representing high heart rate, rather than negative high heart rate), which negatively impacted their ability to interpret mood from the animos.

\paragraph{Limited context} Participants were able to express themselves by creating their own meanings for the animos; however, they did not always converge on those meanings, causing them to miss the sender's intentions. Participants lacked a \textit{shared language} in which to communicate with the animos.

\subsubsection{Recommendations for Content and Presentation} 
\paragraph{Content Provider} To promote expressiveness, designers should consider a ``mixed'' content provider with tools for users to subjectively interpret their biosignals, rather than solely focus efforts towards a system that accurately infers information from biosignals. Users could identify with a message more if they have the ability to define what it means to them, and receivers could likewise recognize that it has meaning for them.

\paragraph{Presentation Abstraction} 
High abstraction in presentation can encourage playfulness and expressiveness, but can be difficult to understand. For lightweight communication, simplifying the presentation is important. We recommend minimizing the number of changes a user needs to attend to. Adding small context cues could help distinguish between different meanings. For instance, participants suggested including short captions, such as ``hi'' or ``good morning'' (P14), as well as using activity- and emotion-specific animos (P24) to distinguish between biosignals affected by physical or emotional stimuli. We also recommend exploring non-visual language. For example, a single tap could represent a ``hi,'' while a long press or double-tap could represent a ``heavier'' event that requires response, such as something stressful.

\paragraph{Presentation Granularity} An animo represents the mood of an individual; thus, participants needed to convey how their animo reflected their mood to their partner. To encourage the development of meanings shared between partners, we recommend exploring \textit{communication-based}\footnote{Unlike in~\cite{buschek2018personal}, we use ``communication'' instead of ``conversation'' since conversation may not necessarily occur through the smartwatch.} presentation granularity, to represent how communication develops between two users. For instance, participants suggested exploring playful ways to highlight that users are connected through the system and \textit{``convey a sense of togetherness''} (P9). This could be through animos interacting with each other (e.g., dancing) to show how users' moods could interact. Other examples could include enhancing messages according to communication frequency, or leaving behind ``gifts'' that serve as a reminder of communication that occurred.

\subsection{Sharing: ``Being There''}

Animo used the smartwatch as a \textit{sharing} channel, and allowed participants to directly share their heart rate with each other. This enabled a new way for people to connect by physically sharing information from their body. Our results suggest that this form of sharing enhanced social presence between participants.

\subsubsection{Social Presence}
Social presence has various definitions, but for our purposes, we adopt Biocca and colleagues' succinct definition: ``the sense of being with another''~\cite{biocca2003toward}. Factors such as a lack of \textit{immediacy} and \textit{intimacy} can reduce social presence when people communicate through mediated channels~\cite{gunawardena1997social,gunawardena1995social}. Promoting social presence should promote feelings of connectedness~\cite{ijsselsteijn2003staying}.

Past research shows that even with minimal communication content, people can feel as if the other person is present~\cite{kaye2005communicating,chang2001lumitouch}. Participants felt that their partners were ``there'' on their wrist, especially when they were not physically together. This stemmed from several factors:

\paragraph{Mediated touch} The smartwatch is worn on the wrist. When a user receives an animo, they can physically feel it through haptic feedback. Participants mentioned using Animo as a ``poke''; haptic feedback can make them feel like they are being poked. Our results suggest the haptic features of the smartwatch in communication can act as a ``mediated social touch,'' which can increase social presence~\cite{van2015social,beelen2013art}.

\paragraph{Immediacy} Immediacy is already valued in smartwatches: users enjoy notifications for their ability to quickly check incoming calls and messages~\cite{jeong2017smartwatch}. Animo leverages this immediacy for two-way communication: participants found that their partners were more immediately accessible to them. They could easily initiate or respond to communication by tapping on the watch to send or recieve an animo, without reaching for their phone or computer, or typing a message out. Animo's form factor and simple design made it ``\textit{effortless}'' (P16) to use, helping participants communicate more frequently, even during busy times.

\paragraph{Intimacy} In addition to the affordances of the smartwatch, Animo's sharing design affords intimate communication between close-ties, such as significant others and close friends. Specifically, Animo is one-to-one. As P11 mentioned, this makes communicating feel \textit{``more personal''} because only one other person can send animos; they act like a \textit{``hidden message''} (P27). 
Moreover, the message sent is ``mood'' derived from heart rate, personal and private information that can increase feelings of intimacy and vulnerability when shared~\cite{janssen2010intimate,liu2017supporting,slovak2012understanding}. Similar to prior work~\cite{hassib2017heartchat,liu2017expressive}, participants noted they would prefer to share Animos with their closest contacts. P25 mentioned that she had only known her partner for a few months, and would have liked to participate with her best friend instead.

\subsubsection{Recommendations for Sharing} We suggest a  new design dimension in biosignals smartwatch communication: \textit{sharing receiver}. Animo allowed only one receiver, which created an intimate communication experience preferred by close partners. For an experience geared towards relationships with varying degrees of closeness, designers might consider multiple receivers for animos, which would allow senders to keep in touch with several people at once.

\subsection{Response: Triggering Conversation}
Similar to prior work, sending and receiving animos helped participants feel more aware of and connected with their partners~\cite{liu2017supporting,hassib2017heartchat}. This supports research suggesting that keeping in touch with a partner's activities can facilitate feelings of connectedness, especially when remote~\cite{lottridge2009sharing,bales2011couplevibe}. At the same time, since Animo was so minimal, it was unable to go beyond these short status updates to support full conversations between partners. Instead, participants used the animos they saw as opportunities to converse with each other through other channels, such as through text when they received animos or face-to-face at later times when they physically met.

Our results suggest that systems like Animo are best used as \textit{conversation starters}, rather than platforms for conversation. For a seamless communication experience, future work should consider how these systems can smoothly transition between a lightweight conversation starter to actual heavier conversation. In particular, we suggest exploring how receivers can better \textit{respond} to biosignals messages to inform or initiate future conversations.

\subsubsection{Recommendations for Response} We recommend \textit{response} as a new design dimension with two sub-dimensions: \textit{response richness} and \textit{response channel}. The former refers to determining situations in which simple or rich responses are appropriate. Animo only enabled simple responses, where participants could send animos back to acknowledge they received their partner's animo. Richer and more diverse responses in-app could better clarify the intent behind the received animo, and guide follow-up conversation (e.g., if a receiver recognizes a sender's animo as an expression of stress, they can provide support in a phone call). For example, P18 suggested enabling short-form replies:

\begin{quote}
\textit{``Like, if I received an Animo and I could just respond with, like, three or four questions without having to type anything, but it'd just be like, `Sad, smiley, question mark'''
 ---P18}
\end{quote}

\textit{Response channel} refers to the channel through which users can respond to a biosignals message. Animo used a single channel, where participants responded to animos within the app on the smartwatch. Another option is to link the app to other communication channels, such as an in-app call button to link to the phone, or pinned animos that users can send in text conversations as a reference. Having multiple linked communication channels could support smoother entry into full conversations from lightweight systems like Animo.

\section{Limitations}
Our research provides valuable insights for using smartwatches as a communication platform with biosignals as content. However, our work has some limitations.

We focused on understanding how people would use a mood-sharing smartwatch app and the opportunities and challenges that would arise. Therefore, we chose to create Animo as the minimal viable prototype, putting our efforts towards the design of the system, rather than incorporating accurate mood detection. 
Future work should investigate how to incorporate enhanced mood detection. For example, in our work, we represented mood based primarily on physiological arousal, but did not determine whether that mood was positive or negative. One way to incorporate valence is to provide options for users to select valence for their mood representation. Providing manual valence options could affect how people understand and find meaning in their biosignals, and subsequently reveal different communication patterns.

Additionally, the demographics of our sample may have biased our results. While the average age of our participants aligns with those who are more likely to own smartwatches, our sample does not include younger adults or children (e.g., 18 and below). Participants were also self-selected, and more than half of them worked at technology companies (and the rest had a partner who did). They likely have more familiarity with technology and may have shown a greater interest in wearables and biosignals than the general public. Our results may not generalize to younger people or people with less familiarity with technology, sensors, or smartwatches.

\section{Conclusion}
We present Animo, a smartwatch app that enables people to communicate their heart rate, in the form of a colored animated shape, to another person. We detail the design and implementation of Animo, as well as findings from a two-week field study with 17 dyads. Participants connected more through Animo, and became more aware of their partner's presence and state. Participants communicated in new ways with each other, using Animo both as a lightweight ``poke'' and a conversation starter. At the same time, Animo revealed design tensions in sensor-based smartwatch communication apps. While participants enjoyed its simplicity, they had to compensate for the lack of context in follow-up communication. They still desired fuller communication with more information and interaction capabilities. We outline a unique design space for smartwatch-based biosignal communication systems, and provide insights on designing for lightweight social connection within this space, including suggestions for enhancing expressiveness, interpretability, and social presence.

\centering


\section*{Supplementary material}

**Supplemental Table 1**. **Animo Pretest Results**
*Means (Standard Deviations) of Mood and Energy ratings from each round of surveys. Both ratings were based on a scale from 1-5 (bad-good mood, and low-high energy). Mood ratings and energy ratings were compared separately, using a repeated measures ANOVA. Purple highlighted values represent animos that were selected for the final app. Means with different letter superscripts are significantly different at p < 0.05 (with Huynh-Feldt's correction)*

### a) Round 1 (N=20)

| ANIMO | MOOD | ENERGY | KEY | |
|---|---|---|---|---|
| Twirl | 4.75 (0.72)$^a$ | 4.70 (0.73)$^a$ | Positive/High Arousal | |
| Stretch up | 4.05 (1.05)$^{a,b}$ | 4.55 (0.61)$^a$ | Negative/High Arousal | |
| Hit walls | 2.05 (1.40)$^{d,e}$ | 4.45 (0.76)$^a$ | Negative/Low Arousal | |
| Jump forward | 2.65 (1.46)$^{c,d,e}$ | 4.45 (0.61)$^a$ | Positive/Low Arousal | |
| Melt down | 2.00 (1.03)$^d$ | 1.90 (1.12)$^b$ | Neutral | |
| Flop | 2.00 (0.97)$^e$ | 1.80 (0.95)$^b$ | Selected animo | |
| Bounce sway | 2.80 (0.89)$^{c,d}$ | 2.20 (1.36)$^b$ | | |
| Twirl fall | 2.60 (1.00)$^{c,d}$ | 2.15 (1.23)$^b$ | | |
| Swing | 3.25 (0.72)$^{b,c}$ | 2.35 (0.88)$^b$ | | |
| Roll forward | 2.85 (0.82)$^{c,d}$ | 2.35 (0.93)$^b$ | | |

### b) Round 2 (N=20)

| ANIMO | MOOD | ENERGY | KEY | |
|---|---|---|---|---|
| Hammock | 3.05 (1.15)$^{a,c,d}$ | 1.75 (0.55)$^{d,e}$ | Positive/High Arousal | |
| Twirl on ground | 3.60 (0.75)$^{a,c}$ | 3.30 (0.73)$^c$ | Negative/High Arousal | |
| Deep breath | 3.20 (0.83)$^{b,c,d}$ | 1.55 (0.89)$^{d,e}$ | Negative/Low Arousal | |
| Bounce on ground | 2.85 (0.81)$^d$ | 2.20 (0.70)$^d$ | Positive/Low Arousal | |
| Bounce | 4.35 (0.81)$^a$ | 4.45 (0.69)$^{a,b}$ | Neutral | |
| Bounce on walls | 4.20 (1.11)$^{a,b}$ | 4.95 (0.22)$^a$ | Selected animo | |
| Stomp | 2.45 (1.54)$^{c,d,e}$ | 4.65 (0.49)$^{a,b}$ | | |
| Shake and run | 2.15 (1.14)$^{c,d,e}$ | 3.80 (0.95)$^b$ | | |
| Melt flatten | 1.75 (0.91)$^e$ | 1.30 (0.73)$^e$ | | |
| Bounce move | 2.65 (0.88)$^{c,d,e}$ | 2.30 (1.03)$^{c,d,e}$ | | |
| Float | 3.00 (0.92)$^{b,c,d}$ | 1.95 (0.95)$^{d,e}$ | | |

**c) Round 3 (N=20)**

*NOTE: For Round 3, we focused specifically on animo designs that targeted the positive/low arousal quadrant of the valence-arousal circumplex. At this point, we already selected enough animos for the other quadrants, but we found the positive/low arousal quadrant challenging in terms of creating animo designs that performed well. Therefore, we decided to have participants on Amazon Mechanical Turk rate new positive/low animos alongside one neutral animo (which acted as an anchor), and compared the results to animos selected from previous rounds using a one-way ANOVA (results of which are in Supplemental Tables 4 & 5).*

| ANIMO | MOOD | ENERGY | KEY | |
|---|---|---|---|---|
| Bounce in place | 3.35 (1.18)[a,b] | 2.45 (1.10)[a] | Positive/Low Arousal | |
| Balloon | 3.60 (0.82)[a] | 2.65 (1.18)[a] | Neutral | |
| Loop | 3.55 (1.05)[a] | 2.80 (1.47)[a] | Selected animo | |
| Deep breath | 2.50 (0.95)[b] | 1.70 (0.99)[b] | | |
| Side hop | 3.40 (1.14)[a,b] | 2.60 (0.88)[a] | | |

**Supplemental Table 2**. **Animo Pretest Round 3 vs Round 1 & 2 Selected Animos**

*One-way ANOVA post-hoc comparisons between Round 3 animos and Round 1 & 2 selected animos for Mood and Energy ratings. (*) represents a significant difference at $p < 0.05$*

**a) Moods**

| Round 1&2 animos (below) / Round 3 animos (right) | Bounce in place | Balloon | Loop | Side hop | Deep breath |
|---|---|---|---|---|---|
| Twirl | | * | * | * | * |
| Stretch up | | | | | * |
| Bounce | | * | * | * | * |
| Hit walls | | * | * | * | * |
| Jump forward | | * | * | | |
| Shake and run | | * | * | * | * |
| Stomp | | * | * | * | |
| Melt down | | * | * | * | * |
| Flop | | * | * | * | * |
| Melt flatten | | * | * | * | * |
| Swing | | | | | |
| Roll forward | | * | * | | |
| Bounce move | | * | * | | |

**b) Energies**

| *Round 1&2 animos (below) / Round 3 animos (right)* | Bounce in place | Balloon | Loop | Side hop | Deep breath |
|---|---|---|---|---|---|
| Twirl | | * | * | * | * |
| Stretch up | | * | * | * | * |
| Bounce | | * | * | * | * |
| Hit walls | | * | * | * | * |
| Jump forward | | * | * | * | * |
| Shake and run | | * | * | * | * |
| Stomp | | * | * | * | * |
| Melt down | | * | * | * | |
| Flop | | * | * | * | |
| Melt flatten | | * | * | * | |
| Swing | | | | | * |
| Roll forward | | | | | * |
| Bounce move | | | | | * |

# Animo Daily Survey

Please fill out this survey about your experience using the Animo app today before the end of the day today. Thank you!

* Required

1. **Participant ID** *

   _____

2. **Please describe how you used the Animo app today (e.g., what did you notice, what functions did you use).** *

   _____
   _____
   _____
   _____
   _____

3. **Did you send any Animos today?** *
   Mark only one oval.

   ◯ Yes    Skip to question 4.
   ◯ No     Skip to question 9.

Please give one or more examples of Animos you sent, including the following details:

4. **Which Animo(s) did you send?** *

   _____

5. **What was happening at the time you decided to send the Animo(s) (e.g., what were you doing)?** *

   _____
   _____
   _____
   _____
   _____

6. **What do you think the Animo(s) meant?** *

   _____
   _____
   _____
   _____
   _____

7. **Why did you decide to send the Animo(s)?** *

   _____
   _____
   _____
   _____
   _____

8. **How did your study partner react to what you sent?** *

   _____
   _____
   _____
   _____
   _____

9. **Were there any Animos you noticed but did not send?** *
   Mark only one oval.

   ◯ Yes    Skip to question 10.
   ◯ No     Skip to question 11.

10. **Why did you not send the Animo(s) you noticed but did not send?** *

    _____
    _____
    _____
    _____
    _____

11. **Did you receive any Animos from your study partner today?** *
    Mark only one oval.

    ◯ Yes    Skip to question 12.
    ◯ No     Skip to question 15.

    Please give one or more examples of Animos you received, including the following details:

12. **Which Animo(s) did you receive?** *

    _____

13. **What do you think the Animo(s) you received meant?** *

    _____
    _____
    _____
    _____
    _____

14. **How did you react to the Animo(s) you received?** *

    _____

15. **Did you encounter anything unexpected while using the Animo app today? If yes, please describe below.**

    _____
    _____
    _____
    _____
    _____

16. **Do you have any general comments based on your usage of the Animo app today?**

    _____
    _____
    _____
    _____
    _____

## Thank you!

You've completed the daily survey for today. For any concerns or issues, please email us at: [redacted].

You can also fill out the following form if you run into any bugs: [redacted]

Please press SUBMIT to submit your responses.

Exit Interview

**General**
1. What was your experience like overall?
2. What did you think?
3. [If they haven't explained already] If you don't mind, could you explain why you would like to drop out of the study? It's alright if you'd prefer not to explain.

**Animo Changes**
1. Did you look at the Animo during the study?
    a. What did you notice in your Animo?
    b. What kinds of changes in your Animo did you usually see?
    c. What did you think that meant?
2. When you felt a vibration indicating a change in your Animo, did you expect those changes?
3. Were there any times you expected to see a change in your Animo but did not?
4. How do you think the changes in the Animo worked?
    a. What do you think drove the different animations
    b. What do you think drove the different colors

**Sharing Behavior**
1. Did you share your Animo?
2. SHARED Animo:
    - When did you choose to share your Animo?
    - What was that decision like for you?
    - How did you feel about sharing?
    - Did you ever add any other information to what you shared, outside of the watch? For example, did you text, or send images or videos?
        i. Were there any times you felt that you could not fully express something you wanted to say to your partner?
    - Did you get any responses from sharing?
    - What do you think they thought when they got the Animo?
    - Did you talk about it at all?
        i. What did you talk about?
3. DID NOT SHARE
    - If no, why did you not want to share?

**Communication Changes**
4. Did you experience any changes in the way you communicated with your partner?
5. If you don't mind, could you explain why you chose this partner for the study?

**Sharing outside of Study**

1. Were there any situations you would have liked to share your Animo but couldn't?
2. Did you share anything with anyone besides your partner?
3. Did you show anyone your Animo face to face?

**Feedback**
1. Do you have any feedback for us about the Animo app? (Other than connection issues)
2. Any other comments or suggestions?

## Debrief Text

Thank you for your participation in the study. We appreciate your patience and understanding with the issues and bugs in our research prototype, as well as your feedback on how we can improve.

To give you more context about the study, we created this application to explore what it would mean to communicate your "mood," derived from your heart rate, to a partner of your choice, and how that would affect your communication with them. The Animos you saw were animations we created based on a variety of moods, such as stress or relaxation. However, since moods can be subjective, we wanted to leave it up to you to interpret it with your partner.

Do you have any questions about this study?

Thanks again, you're free to go.

# Introduction Questionnaire

* Required

1. **Participant ID** *

   _____

## Questions about You

2. **What kind of phone do you use?** *
   Mark only one oval.

   ◯ Android

   ◯ iPhone

   ◯ Other: _____

3. **What model phone do you use? (e.g., Samsung Galaxy S9)** *

   _____

4. **Of the following communication platforms, select the ones that you've used MOST FREQUENTLY IN THE LAST MONTH to contact other people:** *

Check all that apply.

- [ ] Phone call
- [ ] Texting
- [ ] Facetime
- [ ] Snapchat
- [ ] Email
- [ ] WhatsApp
- [ ] Messenger
- [ ] Instagram
- [ ] Twitter
- [ ] Signal
- [ ] Kik
- [ ] Line
- [ ] WeChat
- [ ] Telegram
- [ ] Slack
- [ ] Google Hangouts / Chat
- [ ] Viver
- [ ] KakaoTalk
- [ ] Skype
- [ ] Other: _____

5. **Which of the following smartwatches and fitness trackers have you owned?** *

Check all that apply.

- [ ] Apple Watch
- [ ] Fitbit smartwatch
- [ ] Fitbit wristband
- [ ] Fitbit clip
- [ ] Google Wear OS / Android Wear
- [ ] Pebble
- [ ] Misfit
- [ ] I do not own a smartwatch or fitness tracker.
- [ ] Other: _____

6. **How frequently do you use the smartwatches/fitness trackers you selected?**
   Mark only one oval.
   - ◯ Very frequently
   - ◯ Frequently
   - ◯ Occasionally
   - ◯ Rarely
   - ◯ Very rarely
   - ◯ Never
   - ◯ N/A

7. **If you selected any smartwatch/fitness tracker above, for what reason do you use them?**

   _____

8. **What self tracking apps have you used?** *
   Check all that apply.
   - ☐ Strava
   - ☐ MyFitnessPal
   - ☐ Apple Health
   - ☐ Google Fit
   - ☐ Samsung Health
   - ☐ Fitbit App
   - ☐ I have not used any self tracking apps.
   - ☐ Other: _____

9. **What accessories do you wear on a daily basis?** *
   Check all that apply.
   - ☐ Watch
   - ☐ Ring(s)
   - ☐ Necklace(s)
   - ☐ Earrings
   - ☐ Bracelet(s)
   - ☐ Sunglasses
   - ☐ Glasses
   - ☐ I do not wear any accessories on a daily basis.
   - ☐ Other: _____

10. **Please rate the extent to which you think each pair of traits applies to yourself, even if one characteristic applies more strongly than the other.** *

Mark only one oval per row.

|  | Strongly disagree | Disagree | Somewhat disagree | Neither agree nor disagree | Somewhat agree | Agree | Strongly agree |
|---|---|---|---|---|---|---|---|
| Extraverted, enthusiastic | ○ | ○ | ○ | ○ | ○ | ○ | ○ |
| Critical, quarrelsome | ○ | ○ | ○ | ○ | ○ | ○ | ○ |
| Dependable, self-disciplined | ○ | ○ | ○ | ○ | ○ | ○ | ○ |
| Anxious, easily upset | ○ | ○ | ○ | ○ | ○ | ○ | ○ |
| Open to new experiences, complex | ○ | ○ | ○ | ○ | ○ | ○ | ○ |
| Reserved, quiet | ○ | ○ | ○ | ○ | ○ | ○ | ○ |
| Sympathetic, warm | ○ | ○ | ○ | ○ | ○ | ○ | ○ |
| Disorganized, careless | ○ | ○ | ○ | ○ | ○ | ○ | ○ |
| Calm, emotionally stable | ○ | ○ | ○ | ○ | ○ | ○ | ○ |
| Conventional, uncreative | ○ | ○ | ○ | ○ | ○ | ○ | ○ |

## Questions about your relationship with your partner

For all of these questions, "study partner" refers to the person you came with today for the study.

11. **What is the nature of your relationship with your study partner? (select all that apply)** *

Check all that apply.

- [ ] Parent / child
- [ ] Sibling
- [ ] Friend or acquaintance
- [ ] Coworker
- [ ] Roommate
- [ ] Significant Other / Girlfriend / Boyfriend
- [ ] Spouse
- [ ] Other: _____

12. **How long have you known your study partner?** *

    Mark only one oval.

    - ◯ Less than 1 year
    - ◯ 1 year
    - ◯ 2 years
    - ◯ 3 years
    - ◯ 4 years
    - ◯ 5+ years
    - ◯ Other: _____

13. **If you and your study partner are married, how long have you been married?**

    Mark only one oval.

    - ◯ Less than 1 year
    - ◯ 1 year
    - ◯ 2 years
    - ◯ 3 years
    - ◯ 4 years
    - ◯ 5+ years
    - ◯ Other: _____

14. **If you and your study partner are living together, how long have you been living together?**

    Mark only one oval.

    - ◯ Less than 1 year
    - ◯ 1 year
    - ◯ 2 years
    - ◯ 3 years
    - ◯ 4 years
    - ◯ 5+ years
    - ◯ Other: _____

15. **In the past week, how often have you talked to your study partner face-to-face (in person)?** *

    Mark only one oval.

    - ◯ I did not talk to my partner face-to-face
    - ◯ Once every couple of days
    - ◯ 1-3 times every day or almost every day
    - ◯ More than 3 times every day or almost every day
    - ◯ Continuously throughout the week
    - ◯ Other: _____

16. **In the past week, how often have you talked with your study partner remotely, such as through phone calls, text, or other communication platforms?** *

    Mark only one oval.

    - ◯ I did not talk to my partner face-to-face
    - ◯ Once every couple of days
    - ◯ 1-3 times every day or almost every day
    - ◯ More than 3 times every day or almost every day
    - ◯ Continuously throughout the week
    - ◯ Other: _____

17. **In the past month, which communication platform(s) did you use with your study partner?** *

    Check all that apply.

    - ☐ Phone call
    - ☐ Texting
    - ☐ Facetime
    - ☐ Snapchat
    - ☐ Email
    - ☐ WhatsApp
    - ☐ Messenger
    - ☐ Instagram
    - ☐ Twitter
    - ☐ Signal
    - ☐ Kik
    - ☐ Line
    - ☐ WeChat
    - ☐ Telegram
    - ☐ Slack
    - ☐ Google Hangouts / Chat
    - ☐ Viver
    - ☐ KakaoTalk
    - ☐ Skype
    - ☐ Other: _____

18. **If you selected multiple communication platforms, which of these did you use the most to communicate with your study partner?**

    _____

19. **How would you describe your style of communication with your study partner? (e.g., use of different platforms, length of texts, emoji-usage, frequency, etc.)** *

    _____
    _____
    _____
    _____
    _____

20. **How often do you express the following feelings to your study partner:** *
    Mark only one oval per row.

    |  | Never | Rarely | Sometimes | Most of the time | Always |
    |---|---|---|---|---|---|
    | Excited | ○ | ○ | ○ | ○ | ○ |
    | Relaxed | ○ | ○ | ○ | ○ | ○ |
    | Anxiety | ○ | ○ | ○ | ○ | ○ |
    | Indifference | ○ | ○ | ○ | ○ | ○ |
    | Calm | ○ | ○ | ○ | ○ | ○ |
    | Apathy | ○ | ○ | ○ | ○ | ○ |
    | Happy | ○ | ○ | ○ | ○ | ○ |
    | Sad | ○ | ○ | ○ | ○ | ○ |
    | Tired | ○ | ○ | ○ | ○ | ○ |
    | Anger | ○ | ○ | ○ | ○ | ○ |

21. **Please rate the extent to which you agree or disagree with the following statements:** *
Mark only one oval per row.

|  | Strongly disagree | Disagree | Somewhat disagree | Neither agree nor disagree | Somewhat agree | Agree | Strongly agree |
|---|---|---|---|---|---|---|---|
| My relationship with my study partner is close. | ○ | ○ | ○ | ○ | ○ | ○ | ○ |
| When we are apart, I miss my study partner a great deal. | ○ | ○ | ○ | ○ | ○ | ○ | ○ |
| My study partner and I disclose important personal things to each other. | ○ | ○ | ○ | ○ | ○ | ○ | ○ |
| My study partner and I have a strong connection. | ○ | ○ | ○ | ○ | ○ | ○ | ○ |
| My study partner and I want to spend time together. | ○ | ○ | ○ | ○ | ○ | ○ | ○ |
| I'm sure of my relationship with my study partner. | ○ | ○ | ○ | ○ | ○ | ○ | ○ |
| My study partner is a priority in my life. | ○ | ○ | ○ | ○ | ○ | ○ | ○ |
| My study partner and I do a lot of things together. | ○ | ○ | ○ | ○ | ○ | ○ | ○ |
| When I have free time I choose to spend it alone with my study partner. | ○ | ○ | ○ | ○ | ○ | ○ | ○ |
| I think about my study partner a lot. | ○ | ○ | ○ | ○ | ○ | ○ | ○ |
| My relationship with my study partner is important in my life. | ○ | ○ | ○ | ○ | ○ | ○ | ○ |
| I consider my study partner when making important decisions. | ○ | ○ | ○ | ○ | ○ | ○ | ○ |

## Demographics

22. **What is your gender?**

_____

18:36 • F. Liu et al.23. **What is your age (in years)?**

    _____

24. **What is your ethnicity?**
    Check all that apply.

    ☐ White/Caucasian
    ☐ African American
    ☐ Hispanic
    ☐ Asian
    ☐ Native American
    ☐ Pacific Islander
    ☐ Other: _____

Proc. ACM Interact. Mob. Wearable Ubiquitous Technol., Vol. 3, No. 1, Article 18. Publication date: March 2019.

# Animation Survey

* Required

Please answer the following questions about the animation below.

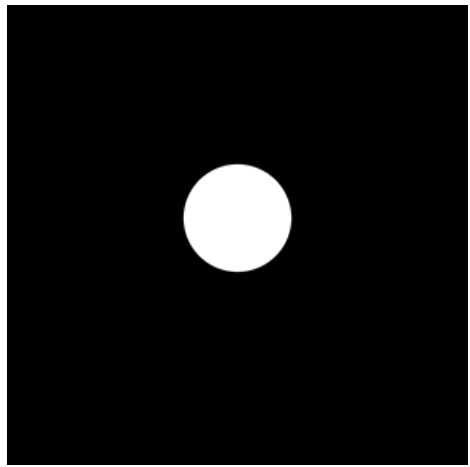

Please use the scale below to indicate the extent to which you believe the circle is in a bad mood versus a good mood. *

|  | 1 | 2 | 3 | 4 | 5 |  |
|---|---|---|---|---|---|---|
| Bad mood | ○ | ○ | ○ | ○ | ○ | Good mood |

How confident are you in your answer above about the circle's mood? *

|  | 1 | 2 | 3 | 4 | 5 |  |
|---|---|---|---|---|---|---|
| Not at all confident | ○ | ○ | ○ | ○ | ○ | Extremely confident |

Please use the scale below to indicate the extent to which you believe the circle is expressing low energy versus high energy. *

|  | 1 | 2 | 3 | 4 | 5 |  |
|---|---|---|---|---|---|---|
| Low energy | ○ | ○ | ○ | ○ | ○ | High energy |

How confident are you in your answer above about the circle's energy? *

|  | 1 | 2 | 3 | 4 | 5 |  |
|---|---|---|---|---|---|---|
| Not at all confident | ○ | ○ | ○ | ○ | ○ | Extremely confident |

What emotion do you think is being expressed by the circle? Please answer to the best of your ability. *

Your answer

Page 2 of 14       BACK       NEXT


\includegraphics[page=17,scale=0.9, trim=55 400 65 40, clip]{supplemental-materials-animo.pdf}
\end{document}